\documentclass[runningheads]{llncs}
\usepackage{graphicx}
\usepackage{style}
\begin{document}
\title{Engineering A Workload-balanced Push-Relabel Algorithm for Massive Graphs on GPUs}
\titlerunning{Workload-balanced Push-relabel Algorithms on GPUs}
%
\author{Chou-Ying Hsieh\and
Po-Chieh Lin\and
Sy-Yen Kuo}
\authorrunning{Hsieh et al.}
%
\institute{National Taiwan University, Taipei, Taiwan \\
\email{\{f07921043, r12921050, sykuo@ntu.edu.tw\}}}


%
\maketitle              
\begin{abstract}
The push-relabel algorithm is an efficient algorithm that solves the maximum flow/ minimum cut problems of its affinity to parallelization. 
As the size of graphs grows exponentially, researchers have used Graphics Processing Units (GPUs) to accelerate the computation of the push-relabel algorithm further.
However, prior works need to handle the significant memory consumption to represent a massive residual graph.
In addition, the nature of their algorithms has inherently imbalanced workload distribution on GPUs.
This paper first identifies the two challenges with the memory and computational models.
Based on the analysis of these models, 
we propose a workload-balanced push-relabel algorithm (WBPR) with two enhanced compressed sparse representations (CSR) and a vertex-centric approach.
The enhanced CSR significantly reduces memory consumption, while the vertex-centric approach alleviates the workload imbalance and improves the utilization of the GPU.
In the experiment, our approach reduces the memory consumption from $O(V^2)$ to $O(V+E)$.
Moreover, we can achieve up to 7.31x and 2.29x runtime speedup compared to the state-of-the-art on real-world graphs in maximum flow and bipartite matching tasks, respectively.
Our code will be open-sourced for further research on accelerating the push-relabel algorithm.

\end{abstract}
\section{Introduction}
The study and application of maximum flow algorithms have long been central to a myriad of computational problems across various disciplines within computer science, 
including VLSI design \cite{qian2016optimal, lyuh2003high}, optimization \cite{rockafellar1999network}, and computer vision \cite{vineet2008cuda}.
The push-relabel (preflow-push) algorithm \cite{goldberg1988new}, in particular, represents a cornerstone in solving the maximum flow problem due to its efficiency and versatility. 
This algorithm iteratively improves the flow within a network by locally pushing excess flow at vertices until the algorithm achieves an optimal flow, leveraging relabel operations to dynamically adjust the heights of nodes to maintain a valid flow.

As graphs grow to encompass billions of nodes and edges,
traditional CPU-based solutions need help with the computational requirements necessary to process such immense datasets efficiently.
This limitation is particularly pronounced in the applications handling large-scale social networks, web graphs, and biological networks, 
where the ability to compute efficiently maximum flow or minimum cut is fundamental but essential for analyzing or understanding graph information. 
The emergence of general-purpose computing on graphics processing units (GPUs) has opened a new road to meet this requirement.
The GPU provides massively parallel processing capabilities, which has significantly reduced the computation time of the push-relabel algorithm in prior works \cite{he2010dynamically, khatri2022scaling}. 
However, those works fail to address the impact of massive graphs.
First, they use an adjacency matrix to represent the graph, 
which takes $O(V^2)$ memory space, where $V$ is the set of vertices.
The enormous memory consumption puts remarkable pressure on a single GPU.
For instance, the most advanced GPU nowadays, H100 NVL with 188 GB VRAM, can only accommodate about 306,594 vertices if we use 2 bytes for a data point in the adjacency matrix. 
Second, the traditional parallel push-relabel algorithm on the GPU has severe workload imbalance on each thread, which fails to utilize the entire computing power of a GPU.

To address these challenges, we propose a novel workload-balanced push-relabel algorithm (WBPR) \footnote{source: https://github.com/NTUDDSNLab/WBPR} designed for modern GPU architectures dealing with massive graphs.
In WBPR, the two enhanced compressed sparse representation (CSR) data structures, RCSR (Reversed CSR) and BCSR (Bidirectional CSR), reduce the significant memory space of a large graph.
Besides, these CSRs can also provide efficient memory access to neighbor vertices for graphs with different characteristics.
Based on these CSRs, we then develop a novel vertex-centric approach of parallel push-relabel algorithms, 
which collects all the active vertices in a queue first, 
so that we can accelerate the local operations of an active vertex by assigning an arbitrary number of threads, 
rather than using a single thread iteratively.
We summarize our contributions as follows:
\begin{itemize}
    \item We derive a computation model to evaluate the execution time of the push-relabel on GPUs. Upon the model, we identify where the weakness and workload imbalance of the state-of-the-art design.
    \item To accommodate massive graphs in a GPU, we proposed RCSR and BCSR, which significantly reduce space complexity from $O(V^2)$ to $O(V+E)$ with trivial overhead on the process of the push-relabel algorithm. We also state that RCSR and BCSR have different advantages in graphs with varying characteristics. 
    \item The novel vertex-centric approach can alleviate the workload imbalance among threads and improve the utilization of the GPU. It achieves an average 2.49x and 7.31x execution time speedup in maximum flow tasks with RCSR and BCSR, respectively; it also gains an average 2.29x and 1.89x execution time speedup in bipartite matching tasks with RCSR and BCSR.
\end{itemize}

We organize the rest of the paper as follows: 
Section \ref{sec:background} introduces the background of the maximum flow problem, the architecture of a GPU, 
and the weaknesses of the traditional approach.
We discuss the BCSR and vertex-centric approach in Section \ref{sec:methodologies}. 
Section \ref{sec:evaluation} presents the evaluation of our design compared to the state-of-the-art one. 
Section \ref{sec:conclusion} concludes.

\begin{algorithm}[H]
\caption{The lock-free push-relabel algorithm}
\label{algo:lockfree_pr}
\DontPrintSemicolon
  
\KwInput{$G(V,E)$: the directed graph, $G_f(V, E_f)$: the residual graph, $c_f(v, u)$: the residual flow on $(u,v)$, $e(v)$: the excess flow of the vertex $v$, $h(v)$: the height of the vertex $v$, $Excess\_total$: the sum of excess flow}
\KwOutput{$e(t)$: the maximum flow value}
\KwData{Initialize $c_f(v, u), e, h$, $Excess\_total \leftarrow 0$}

\tcc{Step 0: Preflow}
\ForEach{$(s, v) \in E$} {
    $c_f(s, v) \leftarrow 0$ \;
    $c_f(v, s) \leftarrow c(s,v)$ \;
    $e(v) \leftarrow c(s,v)$ \;
    $Excess\_total \leftarrow Excess\_total + c(s,v)$ \;
}
\While{$e(s) + e(t) < Excess\_total$} {
    \tcc{Step 1: Push-relabel kernel (GPU)}
    $cycle = |V|$ \;
    \While{$cycle > 0$} {
        \ForEach{$u \in V$ and $e(u) > 0$ and $h(u) < |V|$} {
            $h' \leftarrow \infty$ \;
            \ForEach{$(u, v) \in E_f$} {
                \If{$h' < h(v)$} {
                    $h' \leftarrow h(v)$ \;
                }
            }
            \If{$h(u) > h'$} { 
                $d \leftarrow$ MIN($e(u), c_f(u, v')$$)$ \;
                AtomicSub($c_f(u, v'), d$) \;
                AtomicSub($e(u), d$) \;
                AtomicAdd($c_f(v', u), d$) \;
                AtomicAdd($e(v'), d$)
            }
            \Else{
                $h(u) \leftarrow h' + 1$ \;
            }
        }
        $cycle \leftarrow cycle - 1$ \;
    }
    \tcc{Step 2: Heuristic Optimization (CPU)}
    GlobalRelabel() \;
}
    
\end{algorithm}

\section{Background}
\label{sec:background}

\subsection{The Maximum Flow/ Minimum Cut Problems}
\label{sec:maxflow_def}

Given a directed graph $G(V, E)$, where $V$ and $E$ are the vertex and weighted edge set, respectively, each edge $(u, v)$ has a weight, \emph{capacity} $c(u, v)$, representing the maximum amount of flow that can be passed through this edge. 
The maximum flow problem tries to find the maximum flow value from the source vertex $s$ to the sink vertex $t$. 
Generally, algorithms solve the maximum flow problem operating on the \emph{residual graph } $G_f(V, E_f)$, which has the same vertex set as the given $G$ with \emph{residual edges} $E_f$.
The weight $c_f(u, v)$ of a residual edge $(u, v)$ represents how much-remaining flow can be passed from $u$ to $v$.
Note that when there is a flow from $u$ to $v$, there will be a residual edge from $v$ to $u$, even if there is no $(v,u)$ in $E$.

The foundational algorithm solving the maximum flow problem is the Ford-Fulkerson algorithm \cite{ford1956maximal}, 
which finds augmenting paths from the source to the sink iteratively.
It pushes the additional flow until no such path remains.
Edmonds and Karp improved the efficiency of finding augmenting paths using breadth-first search (BFS). 
The Edmonds-Karp algorithm \cite{edmonds1972theoretical} reduces the runtime complexity of Ford-Fulkerson from $O(Ef)$ to $O(VE^2)$,
where $f$ is the maximum flow value.
Dinic et al. \cite{dinitz2006dinitz} further improved upon the basic augmenting path approach by employing a level graph and blocking flows, 
which significantly reduced the number of augmentations needed;
hence, reducing the complexity to $O(V^2E)$ and $O(V^{3/2}E)$ on general and unit-capacity graph respectively.
Goldberg et al. \cite{goldberg1988new} proposed the push-relabel algorithm (also known as the preflow-push algorithm), which focuses on local operations to adjust the flows.
Some optimizations are based on it to improve the complexity to $O(V^2E)$.

\subsection{Generic Parallel Push-relabel Algorithm}
\label{sec:paraPR}

The basic idea of the push-relabel algorithm is to generate as much flow as possible at the source, 
and gradually push it to the sink.
It introduces the \emph{excess flow} concept that allows a vertex to have multiple flows coming into it; 
namely, it allows a vertex to have more flow coming in than passing out during the push-relabel procedure,
this vertex is called \emph{active}.
We denote $e(v)$ as the excess flow value on vertex $v$.
The procedure of the push-relabel algorithm is to find active vertices and apply \emph{push} and \emph{relabel} operations on them until there are no remaining active vertices.
To find the initial active vertices, 
the algorithm pushes flow from the source to all its neighbor vertices as much as possible, called \emph{preflow}.
A push operation then forces an active vertex to \emph{discharge} its excess flow and pushes to its neighbors in $G_f$.
To avoid the endless pushing, the push-relabel algorithm also introduces the \emph{height function} $h$ on each vertex.
At first, the source height is $|V|$, and the height of vertices except the source is 0.
It forces an active vertex $u$ to push to the vertex $v$ \emph{iff} $h(u) = h(v) + 1$.
If no neighbor vertex satisfies the constraint, 
the active vertex will \emph{relabel} its height by finding the minimum height $h'$ of its neighbor vertices and setting $h(u) \leftarrow h' + 1$. 
With the relabel operation, the height of an active vertex, which cannot push its excess flow at the end, will be increased.
The active vertex will be deactivated once its height exceeds $|V|$.
The whole procedure ends when there are no active vertices.

The push-relabel algorithm has not only the best theoretical and practical complexity,
but it is well-suitable to parallelization due to its inherent structure and the local nature of its operations.
Numerous works have developed the parallel version on the different hardware platforms, such as multiprocessor \cite{anderson1995parallel, hong2008lock}, and graphics processing unit (GPU) \cite{he2010dynamically, wu2012efficient, khatri2022scaling}.
Since there are different trade-offs and optimization used in different algorithms, 
we target the lock-free algorithm proposed by He et al. \cite{he2010dynamically}, 
which is the state-of-the-art algorithm operating on the GPU shown in \ref{algo:lockfree_pr}.
It parallelly checks whether a vertex $u$ is active by assigning a thread in a GPU (Line: 9).
For convenience, we will use thread $u$ to represent the thread, which checks the vertex $u$.
If the vertex is active, the thread $u$ will find its neighbor vertex $v$ whose height is minimum among other neighbor vertices (line:10 - 13).
The thread then push flow from $u$ to $v$ (line:15 - 19) when $h(u) > h(v)$;
otherwise, the thread will relabel the active vertex $u$. 
After \emph{cycles} times of iteration, 
the algorithm applies \emph{global-relabeling} heuristic, one of the push-relabel optimizations, to improve the practical performance \cite{he2010dynamically}.
The global relabeling updates the height of each vertex by performing a backward breadth-first search (BFS) from the sink to the source in the residual graph $G_f$.
The height will be reassigned to the shortest distance from the sink.
After global relabeling, the procedure subtracts \emph{Excess\_total} from the excess flow of those inactive vertices to guarantee the termination of the procedure (line: 6 in Algorithm \ref{algo:lockfree_pr}).
Since the lock-free algorithm assigns a thread to process a vertex, 
we also call this a \emph{thread-centric} approach.

The most significant difference between the original and lock-free push-relabel algorithms is that the lock-free one relaxes the push constraint of the height function $h(v) = h(u) + 1$ to $h(v) > h(u)$ (line: 14).
By finding the minimum-height neighbor and using atomic operations, 
the correctness of the lock-free algorithm has been proven in \cite{hong2008lock}.

\subsection{Execution Model of GPUs}
\label{sec:gpu}

For the convenience of description, 
we use the CUDA (Compute Unified Device Architecture, proposed by Nvidia Corporation) terminologies.
The modern GPU extends SIMD (Single-Instruction-Multiple-Data) to SIMT (Single-Instruction-Multiple-Thread), 
where a warp is the primary computing unit.
A \emph{warp} usually consists of 32 threads, which share a programming counter (PC) and execute in a lockstep manner;
namely, all threads in a warp will execute and access the memory simultaneously; 
hence, the if-else condition inside a warp will cause warp divergence and serialize the execution of a warp, 
leading to tremendous overhead.
On the top of warps, the thread block (TB) executes multiple warps concurrently to hide the latency of memory access on each warp.
Note, all threads in a TB share two fast on-chip memory, \emph{shared memory} and \emph{cache}. 
The former memory is programmable, while the latter is automatic caching.
In addition to this cached memory, the \emph{memory coalescing} technique can also improve memory access efficiency.
If threads inside warp access the memory continuously, 
the hardware will combine these accesses into one request to increase the memory bandwidth.
Multiple TBs comprise an entire grid, which maps to a physical GPU hardware and equips up with \emph{global memory} for data used in a GPU.

\subsection{Workload Imbalance of Push-relabel Algorithm for Large Graph on GPUs}
\label{sec:challenges}

\begin{figure}[t]
    \centering
    \subfigure[]{\includegraphics[width=0.49\textwidth]{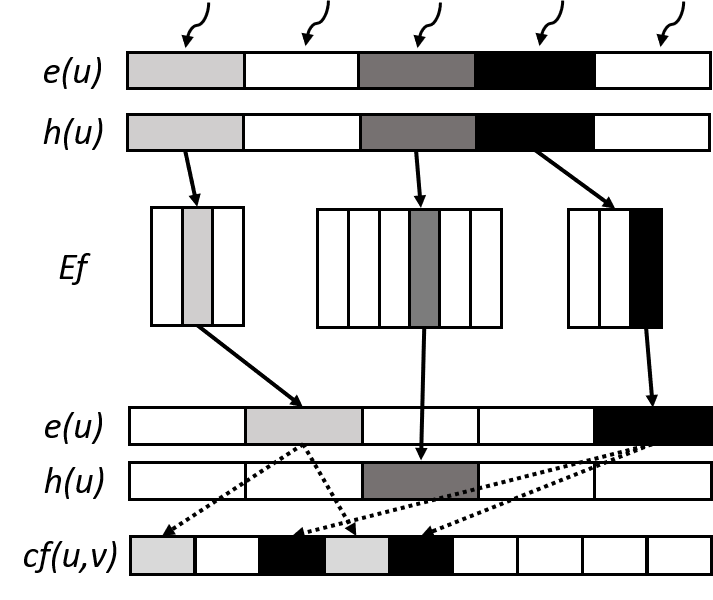}}
    \subfigure[]{\includegraphics[width=0.49\textwidth]{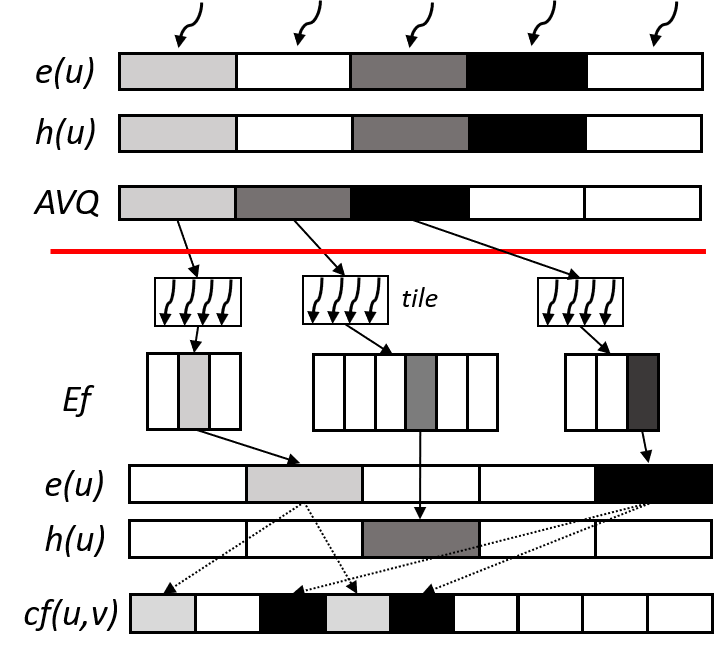}}
    \label{fig:centric}
    \caption{The illustration of the push-relabel algorithm in both (a) thread-centric manner and (b) vertex-centric manner in an iteration. Both two manner approach check the active vertex first ($e(u), h(u)$), and then find the its minimum-height neighbor for push (updating $e(u) $ and $cf(u,v)$) or relabel (updating $h(u)$). The vertex-centric approach uses the AVQ to collect all active vertices, so that it can assign more threads (a tile) for finding a minimum-height neighbor.}
\end{figure}

Despite the parallel lock-free algorithm mentioned in Section \ref{sec:paraPR} has gained excellent performance improvement on GPU, 
the thread-centric approach still faces significant workload imbalance, 
especially when the underlying graph becomes larger and larger.
Specifically, the prior approach fails to fully utilize the parallelism of the GPU. 
Before diving into the imbalance problem, 
we want to derive the cost model of the thread-centric approach on GPU to help us explain where the workload imbalance happens. 
The execution time model can be roughly represented by the following equation:

\begin{equation}
    time = \max\limits_{\{t \in T\}} (\sum_{v}^{V_t}\underbracket[0.8pt]{(k*d(v) + (\lambda_v P(v) + (1-\lambda_v)R(v))}_\text{\clap{execution time of a local operation~}}))
\end{equation}

\noindent 
In this equation, $T$ represents the set of the workers (threads) evolving in this computation, 
and we assume the total active vertex set that the thread $t$ should perform the push and relabel operations on is $V_t$. 
$P(v)$ and $R(v)$ are the time of performing push and relabel on vertex $v$, while the $\lambda_v$, which determines which operation the vertex needs to perform, is 0 or 1.
The $d(v)$ stands for the out-degree of $v$ in the residual graph $G_f$, while $k*d(v)$ represents the total time the vertex $v$ finds its minimum-height neighbor, where $k$ is the constant.
Since the GPU computes in the SIMT manner mentioned in the last section, 
we use the \emph{max} operation to represent the overall execution time, 
which is determined by the last finished thread.
If we want to achieve the optimal overall time, 
we need to divide the workload equally on each thread and reduce the execution time of a local operation in the last finished thread;
namely, let sum of out-degree $d(v)$ in each $V_t$ equally and reduce the time on searching minimum-height neighbor vertex;
However, the prior approach fails to achieve either of the two goals. 
Figure \ref{fig:centric} (a) illustrates the execution of the thread-centric push-relabel algorithm and where the workload imbalance happens.
Since it assigns a thread to check if a vertex is active,
the $V_t$ is not equal.
Besides, the active vertex's out-degree is various, 
so the execution time of the local operation on each active vertex is also various.
Each thread has to iteratively search the incoming and outgoing neighbors in $E_f$ without the help of other threads.
Furthermore, the large graph forces us to use the compressed format, 
such as compressed sparse row (CSR), rather than the adjacency matrix to represent the residual graph.
Finding all outgoing neighbors in the adjacency matrix only takes $O(d(v))$ time complexity. However, it takes up to $O(V*d(v))$ in CSR format, 
exacerbating the impact of workload imbalance.
We will discuss the details in the next section.

\section{Workload-balanced Push-relabel Algorithm}
\label{sec:methodologies}

\subsection{Overview}

Figure \ref{fig:centric} (b) shows the architecture for our workload-balanced push-relabel implementation.
As mentioned in the Section \ref{sec:background}, 
we aim to alleviate the two workload imbalances using the thread-centric approach.
Fundamentally, we first assign all threads to scan all vertices to find the active ones and add the active vertices to the active vertex queue (AVQ).
With the AVQ, each thread has an equal workload when finding active vertices.
Besides, the active vertices can be coalesced in the AVQ; 
hence, we can assign a tile (a group of threads) to find the minimum-height neighbor vertex of an active vertex, 
which reduces the time of the searching time (Algorithm \ref{algo:lockfree_pr} line: 11-13) from $O(d(v))$ to $O(log_2{d(v)})$.
Since we can process multiple vertices and use multiple threads in an active vertex simultaneously, 
we call this approach \emph{two-level parallelism}. 
Furthermore, we designed two enhanced CSRs, \emph{reversed CSR} and \emph{bidirectional CSR}, to reduce significant time in scanning all neighbors or finding backward edges on the residual graph.

\subsection{Enhanced Compressed Sparse Representation}

\begin{figure}[t]
    \centering
    \includegraphics[width=0.95\textwidth]{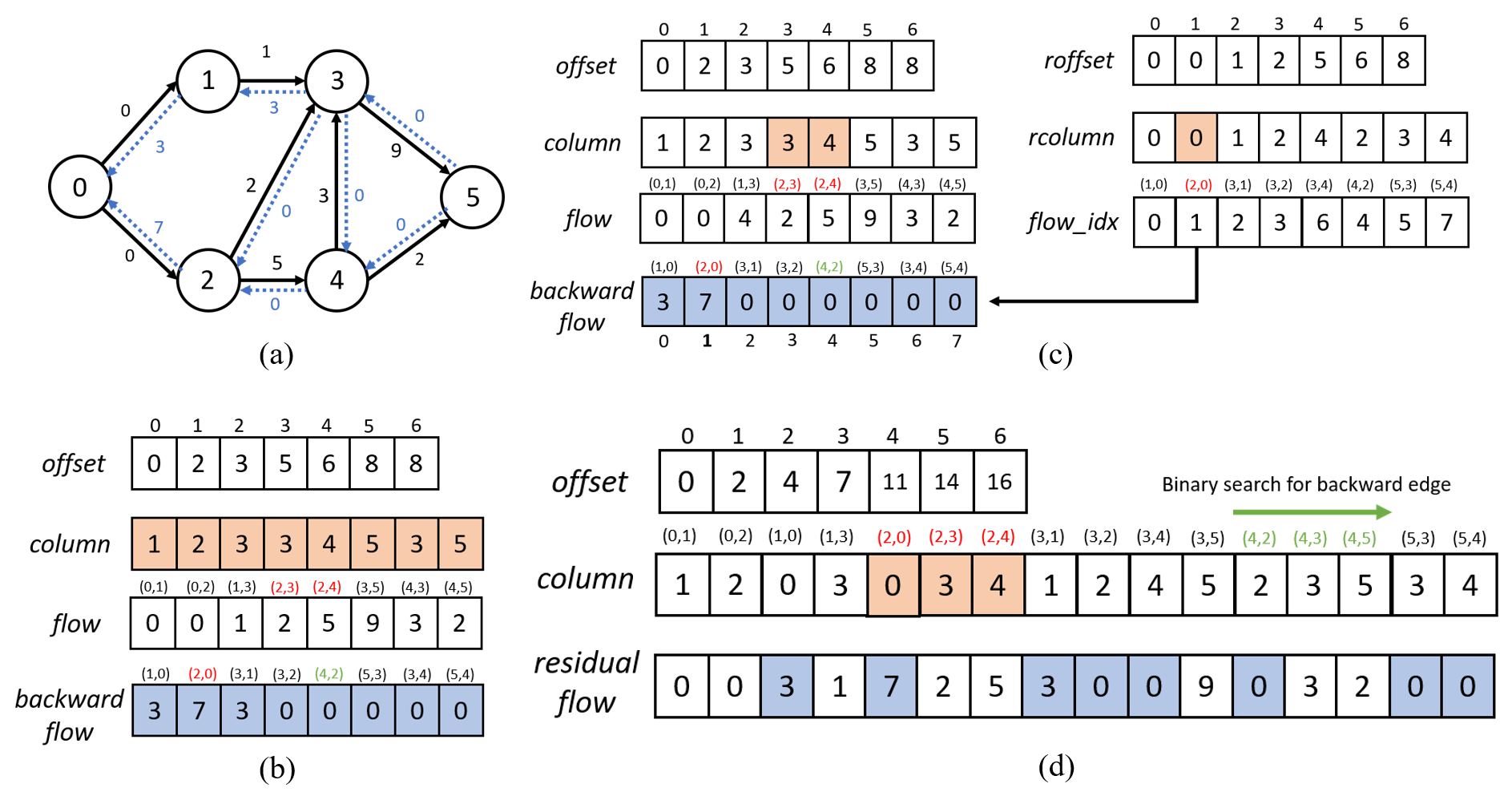}
    \caption{(a) The example residual graph. (b) The original CSR. (c) The reversed CSR (RCSR). (d) The bidirectional CSR (BCSR). The blue block is the flow of backward edges; while the red color stands for all neighbors of vertex $2$ in the residual graph. The orange ones represent the edges to scan when finding the minimum-height neighbor of a given vertex $2$. The green part is the cost to find the backward flow with the given edge $(2, 4)$.}
    \label{fig:csr_combine}
\end{figure}

The prior approach used an adjacency matrix to represent a residual graph, 
so it takes $O(d(u))$ time to find all in- and out-neighbor vertices with a given vertex $u$.
Using an adjacency matrix is convenient and less time-consuming in finding the minimum-height neighbor vertex, 
but it has $O(V^2)$ memory complexity, 
which puts considerable pressure on the GPU.
More and more people have embraced the compressed sparse representation (CSR) to reduce the memory consumption of a massive graph, especially the sparse graph.
However, using a conventional CSR as the residual graph is inefficient for the push-relabel algorithm.
As mentioned in Section \ref{sec:paraPR}, an active vertex must find the minimum-height neighbor $v$ in the residual graph.
After finding the minimum-height neighbor, 
we need to find the backward edge of the $(v, u)$, so that we can decrease $flow(u, v)$ and increase $flow(v, u)$.
If we directly put the backward edge below the forward edge shown in Figure \ref{fig:csr_combine} (b), 
it can access the backward edge in a constant time, 
but it takes $O(|E|)$ to find an active vertex's neighbors.
For instance, finding neighbors of vertex $2$ requires scanning all orange blocks to find $(2, 0)$.

To alleviate the inefficiency of finding incoming neighbor vertices of a given vertex, 
we designed reversed CSR (RCSR) and bidirectional CSR (BCSR) for the residual graph, shown in Figure \ref{fig:csr_combine} (b), (c), respectively.
The RCSR uses another CSR for backward edges.
The \emph{flow\_idx} records the index of backward flow rather than the value.
In this way, we can find all vertex 2's neighbors by scanning the original CSR and reversed CSR (the orange part of Figure \ref{fig:csr_combine} (b)).
Unfortunately, we found that accessing RCSR puts tremendous pressure on the memory bandwidth since neighbors of a vertex are stored in discontinuous addresses, causing uncoalesced memory access.
 
We aggregated incoming and outgoing neighbors to achieve better locality and proposed BCSR shown in Figure \ref{fig:csr_combine} (d).
Since the neighbor is aggregated, 
we cannot access the backward edge in a constant amount of time.
For instance, the edge $(4, 2)$ belongs to vertex $4$'s neighbor; 
Hence, we need to do an additional search to find the backward edge of $(2, 4)$.
If we sort the column list in ascending order by vertex ID, 
we can reduce the searching time from $O(d(v))$ to $O(log_2{d(v)})$ with binary search, where $d(v)$ is the degree of a given vertex $v$.
We evaluate the performance of both RCSR and BCSR in Section \ref{sec:evaluation}.

\subsection{Two-level Parallelism with Vertex-centric Approach}

\begin{algorithm}[H]
\caption{Two-level Parallelism in An Iteration of Push-relabel Kernel}
\label{algo:our_method}

\KwData{$avq$: active vertex queue}

\tcc{Scan the active vertices}
\ForEach{$u \in V$} {
    \If{$e(u) > 0$ and $h(u) < |V|$} {
        $pos \leftarrow$ atomic\_add($avq$, $1$) \;
        $avq[pos] \leftarrow u$ \;
    }
}
grid\_sync() \;

\tcc{First level parallelism}
\ForEach{$u \in avq$} {
    \tcc{Second level parallelism}
    \ForEach{$v \in D(u)$} {
        $min$ = ParallelReduction() \;
    }
    tile.sync() \;
    \If{$localIdx == 0$} {
        \If{$h(v) < h(min)$} {
            Push() \;
        } 
        \Else{
            Relabel() \;
        }
    }
}
\end{algorithm}

Algorithm \ref{algo:our_method} shows the two-level parallelism algorithm for an iteration in Algorithm.\ref{algo:lockfree_pr} (line: 9-21).
In each iteration, we first use the \text{atomic\_add()} operation to add the active vertex to the AVQ.
Since we put a global synchronization in Line:5 (grid\_sync()), 
we can re-organize the thread assignment for the search of minimum-height neighbor.
Besides, we can early break the \emph{while} loop (Algorithm \ref{algo:lockfree_pr}, line: 8) when there is no active vertex remaining in the AVQ, 
which avoids the redundant iteration.
We use a warp as a tile for an active vertex to parallelize the search of a minimum-height neighbor vertex.
Since the neighbor of an active vertex stored in the CSR is continuous, 
using a warp can accelerate the process and reduce the memory request with inherent memory coalescing mentioned in Section \ref{sec:gpu}.
We use the parallel reduction proposed in \cite{harris2007optimizing} to find the vertex with minimum height. 
We chose the Kernel 7 implementation in \cite{harris2007optimizing} since it has the best bandwidth speedup among the seven kernels.
Because the finding minimum operation has a very low arithmetic intensity, 
the peak bandwidth grain can benefit our implementation the most.
After finding the minimum-height neighbor, 
we will use the delegated thread in a warp, whose \emph{localIdx} is 0, 
to execute the push or relabel operations.

\section{Evaluation}
\label{sec:evaluation}

\begin{table}[t]
\caption{The execution time and speedup of different algorithms across 13 graphs. The R0-R10 graphs are the real-world network from SNAP\cite{leskovec2016snap}, while the S0-S1 are the synthesis network generated from 1st DIMACS Challenge \cite{johnson1993network}. The edge capacity of graphs in SNAP is set to 1. The bold font time represents the best execution time among these four algorithms.}
\resizebox{0.99\textwidth}{!}{%
\begin{tabular}{|l|r|r|rrrr|rr|}
\hline
\rowcolor[HTML]{EFEFEF} 
\multicolumn{1}{|c|}{\cellcolor[HTML]{EFEFEF}}                        & \multicolumn{1}{c|}{\cellcolor[HTML]{EFEFEF}}                      & \multicolumn{1}{c|}{\cellcolor[HTML]{EFEFEF}}                      & \multicolumn{4}{c|}{\cellcolor[HTML]{EFEFEF}Execution time (ms)}                                                                                                                                                          & \multicolumn{2}{c|}{\cellcolor[HTML]{EFEFEF}Speedup (TC/VC)}                                          \\ \cline{4-9} 
\rowcolor[HTML]{EFEFEF} 
\multicolumn{1}{|c|}{\multirow{-2}{*}{\cellcolor[HTML]{EFEFEF}Graph}} & \multicolumn{1}{c|}{\multirow{-2}{*}{\cellcolor[HTML]{EFEFEF}$|V|$}} & \multicolumn{1}{c|}{\multirow{-2}{*}{\cellcolor[HTML]{EFEFEF}$|E|$}} & \multicolumn{1}{c|}{\cellcolor[HTML]{EFEFEF}TC+RCSR} & \multicolumn{1}{c|}{\cellcolor[HTML]{EFEFEF}TC+BCSR} & \multicolumn{1}{c|}{\cellcolor[HTML]{EFEFEF}VC+RCSR} & \multicolumn{1}{l|}{\cellcolor[HTML]{EFEFEF}VC+BCSR} & \multicolumn{1}{c|}{\cellcolor[HTML]{EFEFEF}RCSR} & \multicolumn{1}{c|}{\cellcolor[HTML]{EFEFEF}BCSR} \\ \hline
{\color[HTML]{000000} Amazon0302 (R0)}                                & 262,111                                                            & 1,234,877                                                          & \multicolumn{1}{r|}{5,728}                           & \multicolumn{1}{r|}{\textbf{2,307}}                  & \multicolumn{1}{r|}{8,477}                           & 5,191                                                & \multicolumn{1}{r|}{0.67x}                        & 0.44x                                             \\ \hline
{\color[HTML]{000000} roadNet-CA (R1)}                                & 1,965,206                                                          & 2,766,607                                                          & \multicolumn{1}{r|}{57,966}                          & \multicolumn{1}{r|}{70,468}                          & \multicolumn{1}{r|}{74,707}                          & \textbf{32,842}                                      & \multicolumn{1}{r|}{0.78x}                        & 2.15x                                             \\ \hline
{\color[HTML]{000000} roadNet-PA (R2)}                                & 1,088,092                                                          & 1,541,898                                                          & \multicolumn{1}{r|}{27,667}                          & \multicolumn{1}{r|}{\textbf{14,822}}                 & \multicolumn{1}{r|}{43,283}                          & 18,078                                               & \multicolumn{1}{r|}{0.64x}                        & 0.82x                                             \\ \hline
{\color[HTML]{000000} web-BerkStan (R3)}                              & 685,230                                                            & 7,600,595                                                          & \multicolumn{1}{r|}{82,984}                          & \multicolumn{1}{r|}{23,959}                          & \multicolumn{1}{r|}{35,129}                          & \textbf{23,596}                                      & \multicolumn{1}{r|}{2.36x}                        & 1.02x                                             \\ \hline
{\color[HTML]{000000} web-Google (R4)}                                & 875,713                                                            & 5,105,039                                                          & \multicolumn{1}{r|}{28,053}                          & \multicolumn{1}{r|}{12,165}                          & \multicolumn{1}{r|}{17,664}                          & \textbf{7,927}                                       & \multicolumn{1}{r|}{1.58x}                        & 1.53x                                             \\ \hline
{\color[HTML]{000000} cit-Patents (R5)}                               & 3,774,768                                                          & 16,518,948                                                         & \multicolumn{1}{r|}{80,223}                          & \multicolumn{1}{r|}{237,968}                         & \multicolumn{1}{r|}{4,879}                           & \textbf{2,992}                                       & \multicolumn{1}{r|}{16.44x}                       & 79.53x                                            \\ \hline
cit-HepPh (R6)                                                        & 34,546                                                             & 421,578                                                            & \multicolumn{1}{r|}{651}                             & \multicolumn{1}{r|}{214}                             & \multicolumn{1}{r|}{312}                             & \textbf{141}                                         & \multicolumn{1}{r|}{2.09x}                        & 1.52x                                             \\ \hline
{\color[HTML]{000000} soc-LiveJounal1 (R7)}                           & 4,847,571                                                          & 68,993,773                                                         & \multicolumn{1}{r|}{833,189}                         & \multicolumn{1}{r|}{703,077}                         & \multicolumn{1}{r|}{572,705}                         & \textbf{385,912}                                     & \multicolumn{1}{r|}{1.45x}                        & 1.82x                                             \\ \hline
{\color[HTML]{000000} soc-Pokec (R8)}                                 & 81,306                                                             & 1,768,149                                                          & \multicolumn{1}{r|}{82,525}                          & \multicolumn{1}{r|}{66,319}                          & \multicolumn{1}{r|}{73,060}                          & \textbf{34,214}                                      & \multicolumn{1}{r|}{1.13x}                        & 1.94x                                             \\ \hline
{\color[HTML]{000000} com-YouTube (R9)}                               & 1,134,890                                                          & 2,987,624                                                          & \multicolumn{1}{r|}{398,794}                         & \multicolumn{1}{r|}{\textbf{91,859}}                 & \multicolumn{1}{r|}{177,555}                         & 114,003                                              & \multicolumn{1}{r|}{2.25x}                        & 0.81x                                             \\ \hline
{\color[HTML]{000000} com-Orkut (R10)}                                & 3,072,441                                                          & 117,185,083                                                        & \multicolumn{1}{r|}{5,001,263}                       & \multicolumn{1}{r|}{326,534}                         & \multicolumn{1}{r|}{3,141,124}                       & \textbf{325,351}                                     & \multicolumn{1}{r|}{1.59x}                        & 1.00x                                             \\ \hline
Washington-RLG (S0)                                                   & 262,146                                                            & 785,920                                                            & \multicolumn{1}{r|}{162,792}                         & \multicolumn{1}{r|}{132,482}                         & \multicolumn{1}{r|}{287,390}                         & \textbf{99,410}                                      & \multicolumn{1}{r|}{0.56x}                        & 1.33x                                             \\ \hline
Genrmf (S1)                                                           & 2,097,152                                                          & 10,403,840                                                         & \multicolumn{1}{r|}{\textbf{2,138}}                  & \multicolumn{1}{r|}{2,900}                           & \multicolumn{1}{r|}{2,685}                           & 2,503                                                & \multicolumn{1}{r|}{0.79x}                        & 1.15x                                             \\ \hline
\end{tabular}%
}
\label{tab:max_perf}
\end{table}

\subsection{Experiment Setup}

\noindent
\textbf{Applications and datasets.} We evaluated our workload-balanced push-relabel algorithm on both the \emph{maximum flow/ minimum cut} and the \emph{bipartite matching problems}.
We used Washington and Genrmf synthesized networks from the DIMACS 1st Implementation Challenge \cite{johnson1993network} and 10 real-world networks from SNAP \cite{leskovec2016snap} for the maximum flow/ minimum cut problem, 
while we selected 12 real-world bipartite graphs from KONECT \cite{kunegis2013konect} for bipartite matching problem.
Since the real-world networks from SNAP have no specified source and sink vertices, 
we previously used breadth-first-search to find 20 pairs of distinct source and sink vertices with the top 25\% longest diameters.
We add a \emph{super-source} and a \emph{super-sink} connecting to all the 20 sources and sinks, respectively, to compute the multi-source multi-sink maximum flow problem.
In the bipartite matching, the super-source and super-sink connect to two groups of vertices, respectively.
The detailed pair information can be found in our open-sourced repository.

\noindent
\textbf{Implemented algorithms.} Since the prior work \cite{he2010dynamically} did not specify which graph representation it used, we only measured our workload-balanced algorithm with RCSR and BCSR. We implemented the below algorithms:

\begin{itemize}
    \item \textit{\underline{TC+RCSR.}} The thread-centric approach with the RCSR.

    \item \textit{\underline{TC+BCSR.}} The thread-centric approach with the BCSR.

    \item \textit{\underline{VC+RCSR.}} Our vertex-centric approach with the RCSR. 

    \item \textit{\underline{VC+BCSR.}} Our vertex-centric approach with the BCSR. 
\end{itemize}

\noindent
\textbf{Measuring machine.} We ran all the experiments on the Intel i9-10900k 20-core processor @ 3.7GHz with 128GB DDR4 RAM and an Nvidia RTX 3090 GPU.
The number of blocks and block size of the kernel configuration are 1024 and 82, respectively.

\subsection{Performance}

\begin{table}[t]
\caption{The execution time of different push-relabel algorithms across 13 graphs for bipartite matching. The bold font represents the best execution time among the three designs.}
\resizebox{0.95\textwidth}{!}{%
\begin{tabular}{|l|r|r|r|r|rrrr|rr|}
\hline
\rowcolor[HTML]{EFEFEF} 
\multicolumn{1}{|c|}{\cellcolor[HTML]{EFEFEF}}                        & \multicolumn{1}{c|}{\cellcolor[HTML]{EFEFEF}}                      & \multicolumn{1}{c|}{\cellcolor[HTML]{EFEFEF}}                      & \multicolumn{1}{c|}{\cellcolor[HTML]{EFEFEF}}                      & \multicolumn{1}{c|}{\cellcolor[HTML]{EFEFEF}}                                                                         & \multicolumn{4}{c|}{\cellcolor[HTML]{EFEFEF}Execution Time (ms)}                                                                                                                                                          & \multicolumn{2}{c|}{\cellcolor[HTML]{EFEFEF}Speedup (TC/VC)}                                                                                                                                    \\ \cline{6-11} 
\rowcolor[HTML]{EFEFEF} 
\multicolumn{1}{|c|}{\multirow{-2}{*}{\cellcolor[HTML]{EFEFEF}Graph}} & \multicolumn{1}{c|}{\multirow{-2}{*}{\cellcolor[HTML]{EFEFEF}$|L|$}} & \multicolumn{1}{c|}{\multirow{-2}{*}{\cellcolor[HTML]{EFEFEF}$|R|$}} & \multicolumn{1}{c|}{\multirow{-2}{*}{\cellcolor[HTML]{EFEFEF}$|E|$}} & \multicolumn{1}{c|}{\multirow{-2}{*}{\cellcolor[HTML]{EFEFEF}\begin{tabular}[c]{@{}c@{}}Maximum\\ Flow\end{tabular}}} & \multicolumn{1}{c|}{\cellcolor[HTML]{EFEFEF}TC+RCSR} & \multicolumn{1}{l|}{\cellcolor[HTML]{EFEFEF}TC+BCSR} & \multicolumn{1}{c|}{\cellcolor[HTML]{EFEFEF}VC+RCSR} & \multicolumn{1}{c|}{\cellcolor[HTML]{EFEFEF}VC+BCSR} & \multicolumn{1}{c|}{\cellcolor[HTML]{EFEFEF}\begin{tabular}[c]{@{}c@{}}on\\ RCSR\end{tabular}} & \multicolumn{1}{c|}{\cellcolor[HTML]{EFEFEF}\begin{tabular}[c]{@{}c@{}}on\\ BCSR\end{tabular}} \\ \hline
corporate-leadership (B0)                                             & 24                                                                 & 20                                                                 & 99                                                                 & 20                                                                                                                    & \multicolumn{1}{r|}{\textbf{0.15}}                   & \multicolumn{1}{r|}{1.10}                            & \multicolumn{1}{r|}{0.18}                            & 1.00                                                 & \multicolumn{1}{r|}{0.83x}                                                                     & 1.1x                                                                                           \\ \hline
Unicode (B1)                                                          & 614                                                                & 254                                                                & 1,255                                                              & 188                                                                                                                   & \multicolumn{1}{r|}{14.31}                           & \multicolumn{1}{r|}{13.92}                           & \multicolumn{1}{r|}{\textbf{9.12}}                   & 10.84                                                & \multicolumn{1}{r|}{1.57x}                                                                     & 1.28x                                                                                          \\ \hline
UCforum (B2)                                                          & 899                                                                & 522                                                                & 7,089                                                              & 516                                                                                                                   & \multicolumn{1}{r|}{28.87}                           & \multicolumn{1}{r|}{28}                              & \multicolumn{1}{r|}{\textbf{17.04}}                  & 19                                                   & \multicolumn{1}{r|}{1.69x}                                                                     & 1.47x                                                                                          \\ \hline
movielens-u-i (B3)                                                    & 7,601                                                              & 4,009                                                              & 55,484                                                             & 2,836                                                                                                                 & \multicolumn{1}{r|}{289}                             & \multicolumn{1}{r|}{248}                             & \multicolumn{1}{r|}{\textbf{139}}                    & 169                                                  & \multicolumn{1}{r|}{2.08x}                                                                     & 1.47x                                                                                          \\ \hline
Marvel (B4)                                                           & 12,942                                                             & 6,486                                                              & 96,662                                                             & 5,057                                                                                                                 & \multicolumn{1}{r|}{299}                             & \multicolumn{1}{r|}{331}                             & \multicolumn{1}{r|}{\textbf{197}}                    & 240                                                  & \multicolumn{1}{r|}{1.52x}                                                                     & 1.38x                                                                                          \\ \hline
movielens-u-t (B5)                                                    & 16,528                                                             & 4,009                                                              & 43,760                                                             & 3,258                                                                                                                 & \multicolumn{1}{r|}{602}                             & \multicolumn{1}{r|}{567}                             & \multicolumn{1}{r|}{\textbf{305}}                    & 381                                                  & \multicolumn{1}{r|}{1.97x}                                                                     & 1.49x                                                                                          \\ \hline
movielens-t-i (B6)                                                    & 16,528                                                             & 7,601                                                              & 71,154                                                             & 5,882                                                                                                                 & \multicolumn{1}{r|}{621}                             & \multicolumn{1}{r|}{570}                             & \multicolumn{1}{r|}{\textbf{400}}                    & 422                                                  & \multicolumn{1}{r|}{1.55x}                                                                     & 1.35x                                                                                          \\ \hline
YouTube (B7)                                                          & 94,238                                                             & 30,087                                                             & 293,360                                                            & 25,624                                                                                                                & \multicolumn{1}{r|}{157,941}                         & \multicolumn{1}{r|}{110,626}                         & \multicolumn{1}{r|}{\textbf{35,003}}                 & 36,136                                               & \multicolumn{1}{r|}{4.51x}                                                                     & 3.06x                                                                                          \\ \hline
DBpedia\_locations (B8)                                               & 172,079                                                            & 53,407                                                             & 293,697                                                            & 50,595                                                                                                                & \multicolumn{1}{r|}{488,939}                         & \multicolumn{1}{r|}{417,497}                         & \multicolumn{1}{r|}{\textbf{102,373}}                & 109,098                                              & \multicolumn{1}{r|}{4.78x}                                                                     & 3.83x                                                                                          \\ \hline
BookcCrossing (B9)                                                    & 340,523                                                            & 105,278                                                            & 1,149,739                                                          & 75,444                                                                                                                & \multicolumn{1}{r|}{184,985}                         & \multicolumn{1}{r|}{128,079}                         & \multicolumn{1}{r|}{\textbf{64,337}}                 & 70,531                                               & \multicolumn{1}{r|}{2.88x}                                                                     & 1.82x                                                                                          \\ \hline
stackoverflow (B10)                                                   & 545,195                                                            & 96,678                                                             & 1,301,942                                                          & 90,537                                                                                                                & \multicolumn{1}{r|}{864,168}                         & \multicolumn{1}{r|}{631,918}                         & \multicolumn{1}{r|}{309,294}                         & \textbf{267,487}                                     & \multicolumn{1}{r|}{2.79x}                                                                     & 2.36x                                                                                          \\ \hline
IMDB-actor (B11)                                                      & 896,302                                                            & 303,617                                                            & 3,782,463                                                          & 250,516                                                                                                               & \multicolumn{1}{r|}{427,357}                         & \multicolumn{1}{r|}{335,800}                         & \multicolumn{1}{r|}{162,909}                         & \textbf{160,420}                                     & \multicolumn{1}{r|}{2.63x}                                                                     & 2.09x                                                                                          \\ \hline
DBLP-author (B12)                                                     & 5,624,219                                                          & 1,953,085                                                          & 12,282,059                                                         & 1,952,883                                                                                                             & \multicolumn{1}{r|}{\textbf{337,366}}                & \multicolumn{1}{r|}{1,121,432}                       & \multicolumn{1}{r|}{371,953}                         & 609,259                                              & \multicolumn{1}{r|}{0.91x}                                                                     & 1.84x                                                                                          \\ \hline
\end{tabular}%
}
\label{table:biparitie_exe_time}
\end{table}

We measured the kernel execution time of different algorithms.
Table \ref{tab:max_perf} and Table \ref{table:biparitie_exe_time} show the performance of the algorithms in maximum flow and bipartite matching tasks, respectively.
At first glance, our vertex-centric approach can improve the execution of both RCSR and BCSR in the two tasks.
It gains an average 2.49x and 7.31x execution time speedup in maximum flow tasks on RCSR and BCSR, respectively.
In bipartite matching, our VC approach achieves 2.29x and 1.89x execution time speedup on RCSR and BCSR.
We observed two critical points in these experiments.
First, The VC method is suitable for graphs with a high standard deviation of degree,
and these graphs should be a manageable size, 
or the overhead of synchronization can offset the workload balance benefits of VC.
For example, the B0-B2 graphs are too small, so the speedup is marginal.
The performance degradation of graphs R0-R2, S0, S1, and B12 comes from the characteristics of the graph itself.
The Amazon0302 (R0) is the co-purchased product network on Amazon.com.
If a product $v$ is frequently co-purchased with product $u$, 
the graph contains a directed edge from $i$ to $j$.
In this network, almost all nodes are within the same SCC (Strongly Connected Component), and the degrees of these nodes are very close to each other;
hence, the workload of this network is naturally balanced.
The R1 and R2 are two real-world road networks, so their maximum degree is less than 10.
Since we use a tile (usually more than 32 threads per tile) to find neighbors of an active vertex, 
the active node with a small degree will idle most of the threads within a tile and decrease the utilization of the GPU.
On the contrary, if there are some nodes whose degrees are enormous, 
our vertex-centric approach can effectively alleviate the imbalanced workload, such as R5, B7, and B8.

The second observation is that the BCSR outperforms the RCSR in the maximum flow problem.
Nevertheless, there is a slight degradation in the bipartite matching problem.
This result recalls the notion in Section \ref{sec:methodologies} that the RCSR has constant accessing time of backward edges, 
while the BCSR has a better locality than the RCSR.
In the bipartite matching task, since the average degree of a graph is higher than that in the maximum flow task, 
the RCSR performs well because of its fast access to backward edges without searching. 
On the other hand, the better locality of BCSR can reduce the memory request number by memory coalescing, 
leading to better performance.  
\begin{figure}[t]
    \centering
    \includegraphics[width=0.95\textwidth]{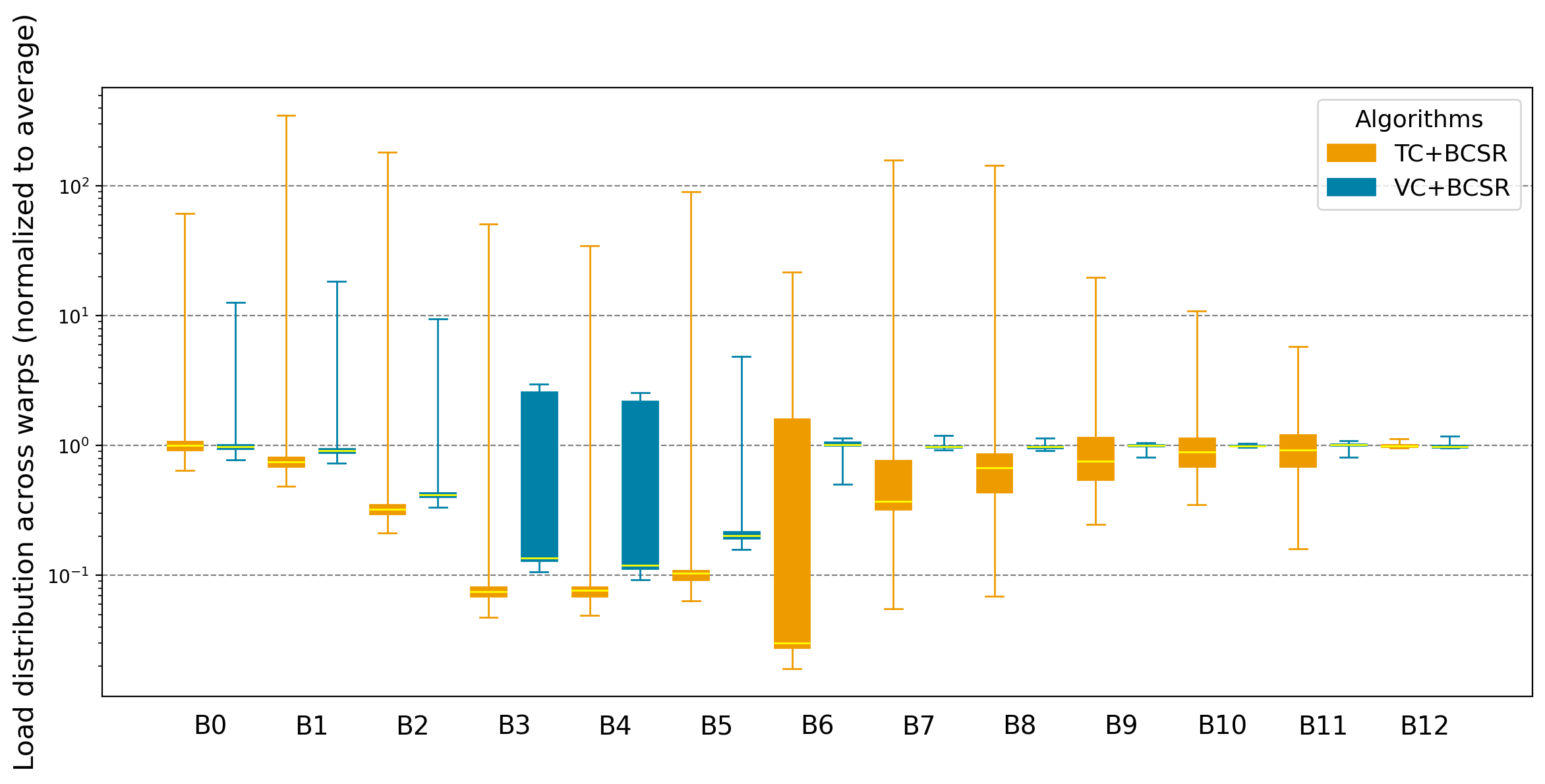}
    \caption{The workload distribution of the bipartite matching problem across 13 bipartite graphs.}
    \label{fig:bipartite_workload}
\end{figure}

\subsection{Workload Analysis}

To prove that the performance improvement of our vertex-centric approach comes from the workload balance during execution, 
we used the method proposed in \cite{almasri2023parallelizing} and measured each warp's execution time.
By assigning the delegated (first) thread within a warp for recording the execution timestamp in the kernel function, 
we can draw the workload distribution for the TC and VC approach across 13 bipartite graphs in Figure \ref{fig:bipartite_workload}. 
Note that both configurations operate on RCSR.
Based on the result, we can make two observations.

First, even though the chart does not directly indicate a shorter execution time for VC (after being normalized by the mean), 
the vertex-centric approach does indeed reduce the standard deviation of execution times across warps, 
thus achieving a more even distribution of work. 

Second, in smaller graphs, even if we equalize the uneven distribution of work among warps, 
the overall performance may still decrease due to excessive synchronization.
The B0, B1, and B2 are the cases.

\section{Related Works}
\label{sec:related_works}

Load balancing is the key factor to achieve high performance when computing graph algorithms on GPUs.
Several fundamental graph algorithms have developed their load balancing heuristics, 
such as SSSP \cite{davidson2014work} (single-source-shortest-path), BFS(breadth-first-search) \cite{hsieh2023decentralized}, triangle counting \cite{hu2019triangle}, clique enumeration \cite{almasri2023parallelizing}, and graph neural network \cite{wang2021gnnadvisor}. 
On the other hand, 
some works simplified the load balancing rule by proposing a unified abstraction and programming model \cite{wang2016gunrock, liu2019simd}.

\section{Conclusion}
\label{sec:conclusion}

In the paper, we first identify the challenges and workload imbalance issue of the traditional parallel push-relabel algorithm with the derived computation model. 
The introduction of enhanced compressed sparse representation data structures, namely RCSR and BCSR, mitigates the memory space challenges posed by large graphs and optimizes memory access patterns for diverse graph characteristics.
Our vertex-centric approach to the parallel push-relabel algorithm further rectifies the issues of workload imbalance and achieves an average 3.45x runtime speedup in maximum flow and bipartite matching tasks.

%
%
%
%
\bibliographystyle{splncs04}
\bibliography{refs}

\end{document}